\UseRawInputEncoding
\documentclass[conference]{IEEEtran}
\usepackage{cite}
\usepackage{amsmath,amssymb,amsfonts}
\usepackage{algorithm}
\usepackage{algorithmic}
\usepackage{graphicx}
\usepackage{textcomp}
\usepackage{xcolor}
\usepackage{url}
\usepackage{hyperref}
\usepackage[all]{hypcap}

\def\BibTeX{{\rm B\kern-.05em{\sc i\kern-.025em b}\kern-.08em
    T\kern-.1667em\lower.7ex\hbox{E}\kern-.125emX}}

\begin{document}

\title{Is Circuit Depth Accurate for Comparing Quantum Circuit Runtimes?}

\author{
    \IEEEauthorblockN{Matthew Tremba, Paul Hovland, and Ji Liu}
    \IEEEauthorblockA{
        \textit{Mathematics and Computer Science Division} \\
        \textit{Argonne National Laboratory} \\
        Lemont, USA \\
        \{mtremba, hovland, ji.liu\}@anl.gov}
}

\maketitle

\begin{abstract}
    Although quantum circuit depth is commonly used to approximate circuit runtimes, it overlooks a prevailing trait of current hardware implementation: different gates have different execution times. Recognizing the potential for discrepancies, we investigate depth's accuracy for comparing runtimes between compiled versions of the same circuit. In particular, we assess the accuracy of traditional and multi-qubit depth for (1) predicting relative differences in runtime and (2) identifying compiled circuit version(s) with the shortest runtime. Finding that circuit depth is not accurate for either task, we introduce a new metric, gate-aware depth, that weights gates' contributions to runtime using an architecture's average gate execution times. Using average gate times allows gate-aware depth to capture variations by gate type without requiring exact knowledge of all gate times, increasing accuracy while maintaining portability across devices of the same architecture. Compared to traditional and multi-qubit depth, gate-aware depth reduces the average relative error of predictions in task (1) by 68 and 18 times and increases the average number of correct identifications in task (2) by 20 and 43 percentage points, respectively. Finally, we provide gate-aware depth weight configurations for current IBM Eagle and Heron architectures.
\end{abstract}


\section{Introduction}

    While quantum algorithms already show potential to outperform their classical counterparts, they continue to face significant challenges from hardware noise~\cite{shor-1997-shor-alg, grover-1996-search, khanal-2023-noise-vq, ladd-2010-qaoa-noise}. In current devices, most noise originates from gate errors or decoherence, in which physical qubits ``drift'' over time to less predictable states~\cite{saki-2021-error-sources}. As a result, the fidelity of computation depends primarily on a given circuit's gate count and runtime. Obtaining accurate values for these characteristics is therefore critical for assessing the fidelity of circuits and, by extension, the performance of the quantum compilation algorithms which optimized them.

    Two primary methods exist for measuring and comparing circuit runtimes. The first, circuit scheduling, operates at the hardware level and constructs an exact timeline and description of the physical process needed to implement the circuit, with the total duration of that process providing the circuit's exact runtime on the device~\cite{alexander-2020-qiskit-pulse, nguyen-2022-xacc-pulse-compilation}. The second method, circuit depth, operates at the circuit level and counts the minimum number of layers a circuit can be partitioned into, each of which can be interpreted as a sequential step during execution~\cite{nielsen-2010-quant-comp}.
    
    While the number of steps gives a loose proxy for the circuit's runtime, their correlation may be inexact because of differences in gate execution times~\cite{chen-2024-ionq-gate-times, acharya-2024-willow-gate-times, pino-2021-quantinuum-gate-times, ibm-qubit-variability}. For circuits of different sizes this effect may be negligible, but for circuits of similar sizes --- and especially different compiled versions of the same circuit --- they may lead to inaccurate runtime comparisons.
    
\begin{figure}[tbp]
        \centerline{\includegraphics[width=0.45\textwidth]{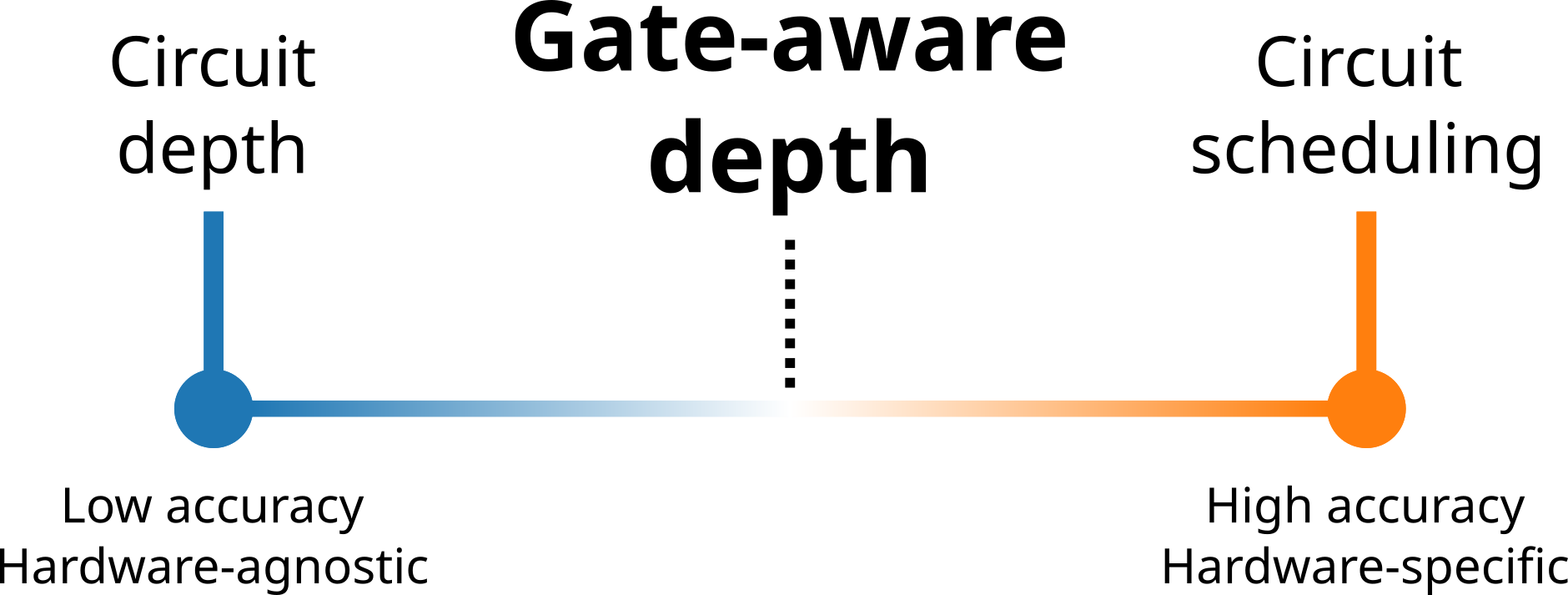}}
        \caption{Gate-aware depth, our proposed new metric, fills the gap in runtime estimation methods, achieving greater accuracy than circuit depth and greater portability than circuit scheduling.}
        \label{fig:spectrum}
    \end{figure}

    As a result, circuit scheduling and circuit depth fall on opposite ends of the accuracy-portability spectrum (see Fig.~\ref{fig:spectrum}). On one hand, circuit scheduling provides perfectly accurate runtime comparisons, but depends on hardware parameters that may vary considerably between devices. On the other hand, circuit depth is completely independent of the target device, but may provide inaccurate runtime comparisons. This raises two questions:
        \begin{enumerate}
            \item Is circuit depth accurate for comparing runtimes, particularly between different compiled versions of the same circuit?
            \item If not, is there an intermediate metric that provides higher accuracy while maintaining portability?
        \end{enumerate}

    In this paper, we address these questions by first proposing a new metric, gate-aware depth, to fill the intermediate role and then assessing the accuracies of both it and the existing depth metrics for circuit runtime comparison. Gate-aware depth weights gates' contributions to circuit runtime using an architecture's average gate execution times, taking advantage of device consistency within an architecture to provide reasonable accuracy across many devices simultaneously. 
    
    We then assess the accuracy of gate-aware, circuit, and multi-qubit depth for comparing runtimes between different compiled versions of the same circuit. In particular, we answer:
        \begin{enumerate}
            \item How accurately do relative differences in the metrics predict relative differences in runtime between compiled circuit versions?
            \item How accurate are the metrics at identifying the compiled circuit version(s) with shortest runtime?
        \end{enumerate}
    
    Our evaluation shows that circuit depth is not accurate for comparing compiled circuit versions, and that gate-aware depth is highly accurate while maintaining the cross-architecture portability that circuit scheduling lacks. Additionally, we provide suitable gate-aware weight configurations for existing IBM Eagle and Heron architectures.

\section{Background}

    \subsection{Quantum Circuit Optimization}
        To mitigate the hardware noise of NISQ-era devices, quantum programs are typically optimized by quantum compilers before they are physically run. A variety of compiler optimization and circuit mapping techniques~\cite{li-2024-qutracer-opt, liu-2025-quclear-opt, liang-2023-hybrid-opt,  campbell-2023-superstaq-opt, liu-2023-qcontext-opt, liu-2021-relaxed-opt, jin-2024-tetris-opt, younis-2021-qfast-opt, nottingham-2023-decomposing-opt, xu-2025-optimizing-opt, xu-2022-quartz-opt, li-2023-sqgm, fosel-2021-quantum-opt, ruiz-2025-quantum-opt, nam-2018-automated-opt,  cao-2024-marqsim-opt, hietala-2021-verified-opt, paykin-2023-pcoast-opt, liu-2023-tackling-opt,  liu-2022-not-opt, tang-2024-alpharouter-opt, molavi-2022-qubit-opt} have been proposed to reduce circuit gate count and depth, since these are indicators for the primary sources of noise~\cite{brugière-2021-reducing-depth}.

        In this context, circuit depth acts as an indicator for runtime, which is itself an approximation of the total decoherence experienced by the underlying quantum system~\cite{ganjam-2024-millisecond-coherence}. While methods like circuit scheduling could provide exact runtime for use as an optimization objective, circuit depth is both simple and widely implemented in compiler frameworks such as Qiskit, TKET, and BQSKit~\cite{qiskit2024, sivarajah-2021-tket, bqskit}.

        Accordingly, researchers assess the runtime optimization capabilities of algorithms via the depths of the circuits they produce~\cite{nation-2025-benchmarking-compilers, zou-2024-lightsabre}. By compiling circuits with multiple algorithms and comparing the relative depths between the different versions of each circuit, researchers can estimate the average-case runtime improvement produced by each algorithm.

    \subsection{Circuit Schedulers}
        Once a quantum program has been optimized in the circuit model, a circuit scheduler compiles the program to the next lowest layer of abstraction by constructing a timeline of instructions that implement that circuit on the underlying quantum system~\cite{nguyen-2022-xacc-pulse-compilation, silverio-2022-pulser, quil-t}. For example, schedulers for IBM's current devices convert hardware-compliant circuits to a collection of microwave pulses that are then applied to manipulate the superconducting qubits~\cite{alexander-2020-qiskit-pulse}.

        Besides its usual applications to tasks like gate calibration and error mitigation~\cite{alexander-2020-qiskit-pulse, werninghaus-2021-leakage-reduction}, circuit scheduling can also be used to measure circuit runtime. Since the schedule specifies the exact physical process implementing the circuit, its total duration is precisely that circuit's quantum runtime. 

    \subsection{Circuit Depth}
        Another option for describing circuit runtime is circuit depth, of which the two main varieties are depth and multi-qubit depth.
        
\begin{figure}[tbp]
            \centerline{\includegraphics[width=0.45\textwidth]{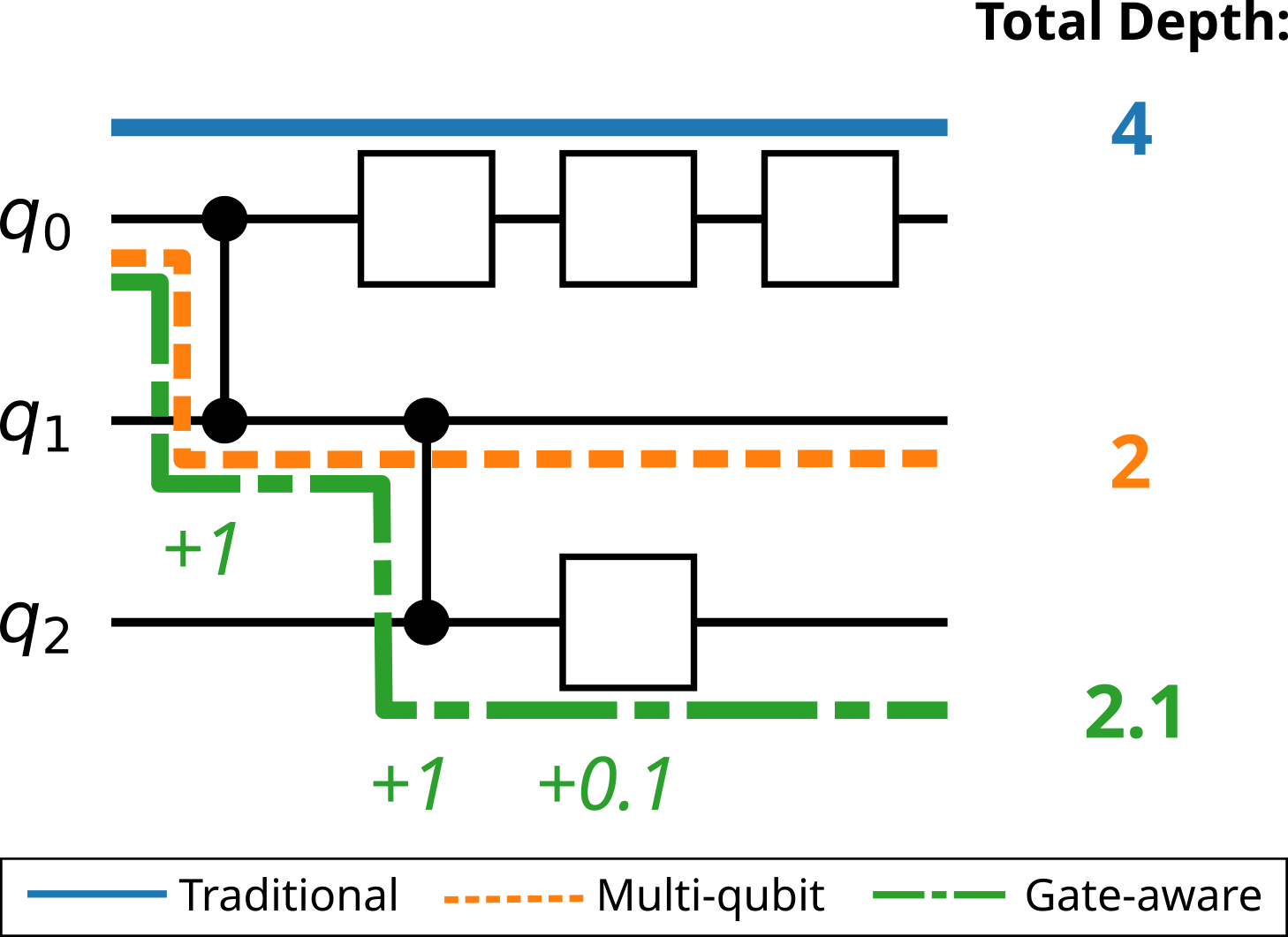}}
            \caption{An example circuit with three possible paths of logically dependent gates, each of which is a critical path for a different metric. Gate-aware depth is configured with weights of 1 and 0.1 for two- and one-qubit gates, respectively.}
            \label{fig:depth-ex}
        \end{figure}

        Depth, referred to hereafter as \emph{traditional depth}, counts the minimum number of layers a circuit can be partitioned into, or equivalently the number of gates in the circuit's longest path of logically dependent gates (the circuit's \emph{critical path})~\cite{nielsen-2010-quant-comp}. For example, the circuit in Fig.~\ref{fig:depth-ex} has a traditional depth of 4 because the critical path, shown in solid blue, contains 4 gates. This path is critical because all other paths, shown in dotted orange and alternating green, have less than 4 gates.
        
        In practice, traditional depth is calculated by updating each qubit's depth while sweeping through the circuit and taking the maximum of the final depths. Each time a gate appears on qubits \( \{q_i, \cdots, q_j\} \), the tracker updates those qubits' depths to
            \begin{equation}
                \label{eq:depth-update}
                d_{\text{new}} = \max_{q_k \in \{q_i, \cdots, q_j\}}(d_{\text{prev}}[q_k]) + 1
            \end{equation}

        Multi-qubit depth, also called entangling depth or CNOT depth, works similarly, but only counts the multi-qubit gates along a given path~\cite{remaud-2025-cnot-depth, debrugière-2024-entangling-depth}. Algorithmically, this can be accomplished by toggling the increment value in \eqref{eq:depth-update} between 0 and 1 for single- or multi-qubit gates, respectively. The circuit in Fig.~\ref{fig:depth-ex} has a multi-qubit depth of 2 because the most multi-qubit gates lying along a path is 2. Note that both the dotted orange and alternating green paths achieve this maximum.

    \subsection{Gate Times}
        For most current quantum devices, individual gate execution times vary with several factors, including number of qubits~\cite{chen-2024-ionq-gate-times, acharya-2024-willow-gate-times, pino-2021-quantinuum-gate-times}, rotation gate angles~\cite{acharya-2024-willow-gate-times}, and the gate's location on the device's physical qubits~\cite{ibm-qubit-variability}. 
        
        While individual gate times vary, average gate times tend to be consistent across devices of similar design. For example, both IBM and IonQ devices implement \( RZ \) gates virtually via phase propagation, meaning these gates do not contribute to circuit runtime on any of these devices~\cite{ibm-virtual-rz, ionq-virtual-rz}. 

    \subsection{Motivation}
        When measuring and comparing circuit runtimes, a metric's accuracy and portability depends on the granularity of gate times it considers. Circuit scheduling and circuit depth represent the two extremes of incorporating every individual gate time or none at all, which results in correspondingly extreme sacrifices in accuracy or portability. Accordingly, a new metric which uses average gate times has the potential to increase accuracy over depth while maintaining portability across the family of devices where those averages reasonably hold.

\section{Gate-Aware Depth}

    Algorithm~\ref{alg:gate-aware-depth} shows the pseudo-code for our proposed metric, \emph{gate-aware depth}. Like traditional depth, gate-aware depth sweeps through the circuit while updating each qubit's depth as gates pass. However, gate-aware depth replaces traditional depth's constant increment with a weight map \( W_{\text{arch}} \) from the native gate set to weights.
    
    \begin{algorithm}[t]
        \caption{Gate-Aware Depth}
        \begin{algorithmic}[1]
        \renewcommand{\algorithmicrequire}{\textbf{Input:}}
        \renewcommand{\algorithmicensure}{\textbf{Output:}}
            \REQUIRE quantum circuit $C$, architecture weight map \( W_{\text{arch}} \)
            \ENSURE gate-aware depth
            \STATE qubit\_depths = [0 for qubit in $C$.qubits]
            \FOR {gate $G$ in $C$}
                \STATE qubits = $G$.location
                \STATE new\_depth = max(qubit\_depths[qubits]) + \( W_{\text{arch}}(G) \)
                \STATE qubit\_depths[qubits] = new\_depth
            \ENDFOR
            \RETURN max(qubit\_depths)
        \end{algorithmic}
        \label{alg:gate-aware-depth}
    \end{algorithm}
    
    The weight map \( W_{\text{arch}} \) is configured using the target architecture's average gate times. For an architecture with native gate set $S$ and gate \( G^* \in S \), we define
        \begin{equation}
            \label{eq:weight-config}
            W_{\text{arch}}(G^*) = \frac{ \text{average gate time of } G^* }{ \max_{G \in S} ( \text{average gate time of } G ) }
        \end{equation}

    Fig.~\ref{fig:depth-ex} depicts the calculation of gate-aware depth for an example circuit. Under the given weight map, the alternating green path has the highest weighted sum of gates, which in turn defines the circuit's gate-aware depth to be 2.1.

    Gate-aware depth bears similarities to the use of cycles per instruction (CPI) for classical processor performance assessment, which incorporates both CPU instruction count and mixture~\cite{hennessy-2011-CPI}.

\section{Experimental Setup}

    Our experiment consists of three phases: compiling a standard circuit test suite for a target device using different algorithms, obtaining the compiled versions' metric values and runtimes, and assessing each metric's accuracy at 1) predicting relative differences in runtime and 2) identifying runtime-optimal circuit versions. We repeated this experiment for IBM's two available superconducting architectures, the 127-qubit Eagle and 156-qubit Heron, using three devices for each architecture.
    
    \subsection{Circuit Test Suite}
        The circuit test suite consisted of 15 real quantum programs from 4 to 64 qubits that are commonly used for compiler benchmarking. We included 4-, 8-, and 16-qubit versions of the VQE~\cite{peruzzo-2014-variational-vqe} and QAOA~\cite{farhi-2014-quantum-QAOA} algorithms, as well as 4-, \mbox{8-,} 16-, and 32-qubit Hamiltonian simulation circuits~\cite{bassman-2020-towards-hamiltonian-sim}. For scale, we additionally included QFT circuits of size 4, 8, 16, 32, and 64. The QAOA, VQE, and Hamiltonian simulation circuits were generated using SupermarQ~\cite{tomesh-2022-supermarq}, while the QFT circuits were generated using Qiskit~\cite{qiskit2024}. 

    \subsection{Compilation}
        The test suite circuits were compiled using four different algorithms: SABRE~\cite{li-2019-sabre}, single-qubit gates matter (SQGM)~\cite{li-2023-sqgm}, Qiskit's default transpiler pass at optimization level 3~\cite{qiskit2024}, and a custom pass in TKET~\cite{sivarajah-2021-tket}. Each circuit was compiled 5 times by each algorithm, and the optimized output with the lowest traditional depth was kept. Since SABRE and SQGM map circuits without rebasing for a device's native gate set, all optimized circuits were subsequently translated for each device using Qiskit's default transpiler pass at optimization level 0 (i.e. no unecessary optimization) to control for the effects of rebasing on depth. 
    
    \subsection{Metric and Runtime Implementation}
        To obtain circuit depths, we implemented gate-aware depth in the BQSKit framework~\cite{bqskit} and used it alongside the pre-existing functions for traditional and multi-qubit depth.
    
        To obtain circuit runtime, we built a custom runtime estimator in BQSKit. The estimator uses Algorithm~\ref{alg:gate-aware-depth}, replacing the average weight map \( W_{\text{arch}} \) with the exact runtimes of all individual gates on the device. With this modification the estimator produces the same runtime as a true circuit scheduler.

        The gate times themselves were accessed through the IBM backends' \verb|instruction_durations| property, which specifies device gate times by both gate operation and qubit location. They are available through the Qiskit IBM runtime package~\cite{qiskit2024}.

        We used the runtime estimator because pulse scheduling was deprecated in Qiskit 1.3, and is not publicly available on the newest Heron architecture. On Eagle devices where the estimator can be verified, its estimates match the pulse scheduler's runtime almost exactly (see Section~\ref{sec:estimator-verification}).

    \subsection{Measures of Accuracy}
        \label{sec:meas-accuracy}
        
        \subsubsection{Predicting Relative Runtime Differences}
            We measure a metric's accuracy for predicting relative differences in runtime between compiled circuit versions using percent relative error (\%RE). For circuits \( C_1 \) and \( C_2 \) which are different compiled versions of the same base circuit, we calculated the relative differences in the metric \( \Delta D \) and runtime \( \Delta R \) using the equations
                \begin{equation}
                    \Delta D = [ \text{depth}(C_1) - \text{depth}(C_2) ] \; / \; \text{depth}(C_2)
                \end{equation}
                \begin{equation}
                    \Delta R = [\text{runtime}(C_1) - \text{runtime}(C_2)] \; / \; \text{runtime}(C_2)
                \end{equation}
                
            Researchers use \( \Delta D \) to predict \( \Delta R \), so we calculate the \%RE of that prediction as
                \begin{equation}
                    \text{\%RE} = | \Delta D - \Delta R | \; / \; | \Delta R | \times 100
                \end{equation}
            This yields the metric's error in predicting the relative difference in runtime between a single pair of compiled circuit versions. By repeating this calculation for many pairs of compiled circuit versions, we create a distribution of errors that captures the metric's overall performance. 

        \subsubsection{Identifying Runtime-Optimal Circuit Versions}
            We measure a metric's accuracy for identifying runtime-optimal circuit versions using the percentage of correct identifications. An identification is correct if, for a given metric, the sets of compiled versions with the minimum metric value and minimum runtime match. This provides the accuracy rate for a common use case in compiler comparison in which the algorithm which minimizes a circuit's depth is presumed to have minimized its runtime too.
    
    \subsection{Platform}
        All tests were conducted with Python 3.13.2 on a 4-core AMD Ryzen 5 3500U with 5.66 GiB main memory running Manjaro 6.6.80. Because SQGM builds on SABRE as implemented in Qiskit version 0.33.0, we used version 0.46.3 of Qiskit Terra for compatibility when compiling with these algorithms. Otherwise, we used versions 0.5.38, 1.4.1, 0.36.1, 1.2.0, and 2.0.1 for the SupermarQ, Qiskit, Qiskit IBM runtime, BQSKit, and PyTKET packages, respectively.

\section{Results}

    All experimental source code and results are available at \url{https://github.com/mtkgv/cdaa}. Additionally, the circuit runtime estimator has been made available as a stand-alone tool at \url{https://github.com/mtkgv/qcre}.

    \subsection{Runtime Estimator Verification}
        \label{sec:estimator-verification}
        
        We verified the accuracy of our custom runtime estimator by comparing it against the Qiskit pulse scheduler. Since only Eagle backends support pulse scheduling, we compiled the 15 test circuits for the IBM Sherbrooke, an Eagle-model device, using SABRE, SQGM, and Qiskit. We then compared our estimated runtime against the pulse schedule's total duration.
        
        For 30 of the 45 compiled circuits, the difference between the estimated runtime and pulse schedule duration was exactly 0 seconds. For the remaining 15 circuits, the largest difference was \( 3.5 \times 10^{-16} \) seconds for a circuit with a true runtime of \( 1.8 \times 10^{-3} \) seconds, which is explainable by floating-point errors. We therefore conclude that our estimated runtime matches true runtime extremely accurately.

    \subsection{Architecture Weight Maps}
        \label{sec:weight-maps}
        
        We configured the gate-aware depth weight maps for our two target architectures, the IBM Eagle and Heron, by taking average gate times over 3 devices each. For the Eagle, we took average gate times over the Sherbrooke, Kyiv, and Brisbane devices; and for the Heron, we took average gate times over the Marrakesh, Kingston, and Aachen devices. The resulting weight maps are shown in Table~\ref{tbl:avg-weight-maps}.
        
        \begin{table}[tbp] 
            \begin{center}
                \caption{Gate-Aware Depth Weight Maps for IBM Eagle and Heron Architectures}
                \label{tbl:avg-weight-maps}
                \begin{tabular}{c||c|c|c|c|c} 
                    $G$ & $ECR$ & $CZ$ & $RZ$ & $SX$ & $X$ \\
                    \hline
                    \( W_{\text{Eagle}}(G) \) & $1.000$ & n/a & $0.000$ & $0.0942$ & $0.0942$ \\
                    \( W_{\text{Heron}}(G) \) & n/a & $1.000$ & $0.000$ & $0.483$ & $0.483$ \\
                \end{tabular}
            \end{center}
        \end{table}

    \subsection{Case Study: QFT64, SQGM v. Qiskit}
\begin{figure}[tbp]
            \centerline{\includegraphics[width=0.5\textwidth]{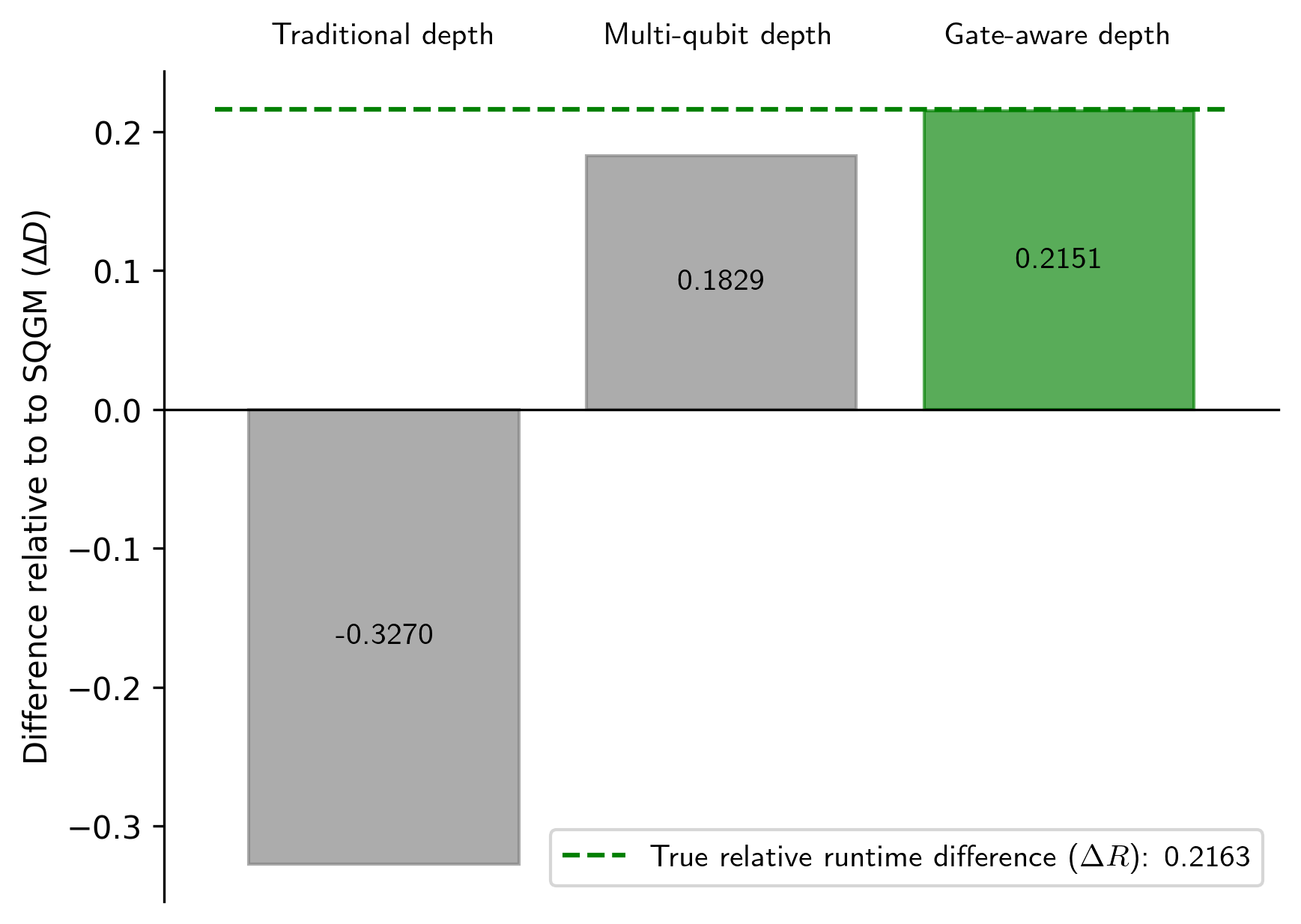}}
            \caption{Proportional change in metrics and runtime of the Qiskit-compiled QFT64 relative to the SQGM-compiled QFT64. The true proportional change in runtime is indicated by the dotted green line.}
            \label{fig:case-study}
        \end{figure}

        Before comparing the metrics' overall accuracies, we first present a single-comparison case study using the 64-qubit QFT circuits compiled by SQGM and Qiskit for the IBM Marrakesh. Fig.~\ref{fig:case-study} illustrates the proportional changes \( \Delta D \) and \( \Delta R \) between these two compiled versions.

        This comparison illustrates a ``worst-case'' scenario for traditional depth because it decreased from the SQGM to Qiskit versions (indicated by negative \( \Delta D \) value) while runtime increased (indicated by the positive \( \Delta R \) value). In comparison, multi-qubit and gate-aware depth both increased, correctly predicting the increase in runtime. However, out of these two metrics, gate-aware depth's relative difference comes closest in magnitude to the relative difference in runtime, making it the most accurate.

        The \%RE outlined in Section~\ref{sec:meas-accuracy} captures these observations numerically; \%RE is higher when \( \Delta D \) and \( \Delta R \) are far apart, and in particular is at least \( 100\% \) when their signs differ. When comparing this pair of circuit versions, traditional, multi-qubit, and gate-aware depth have \%REs of \( 251\% \), \( 15\% \), and \( 0.6\% \), respectively.
    
    \subsection{Accuracy for Predicting Relative Differences in Runtime}
        \label{sec:accuracy-rel-diff}
        
        Next, we compare the metrics' overall accuracies for predicting relative differences in runtime by obtaining a distribution of the \%REs for all possible pairwise comparisons. With 15 circuits and 4 compilers, there were \( \binom{4}{2} \times 15 = 90 \) possible pairwise comparisons, and thus \( n=90 \) data points, for each metric on every device. \footnote{For Eagle devices, our custom TKET compilation pass produced circuits that matched the coupling graph and gate set but occasionally reversed the qubit direction of available \(ECR\) gates. The optimization level 0 Qiskit translation pass raised errors when encountering these reversed \(ECR\)s, preventing us from continuing our experimental procedure. For this reason, we excluded TKET-compiled circuits from our Eagle analyses, giving only \( \binom{3}{2} \times 15 = 45 \) pairwise comparisons for these devices. However, the trends observed remained similar to Heron devices, where all four compilers were included.}
        
\begin{figure}[tbp]
            \centerline{\includegraphics[width=0.5\textwidth]{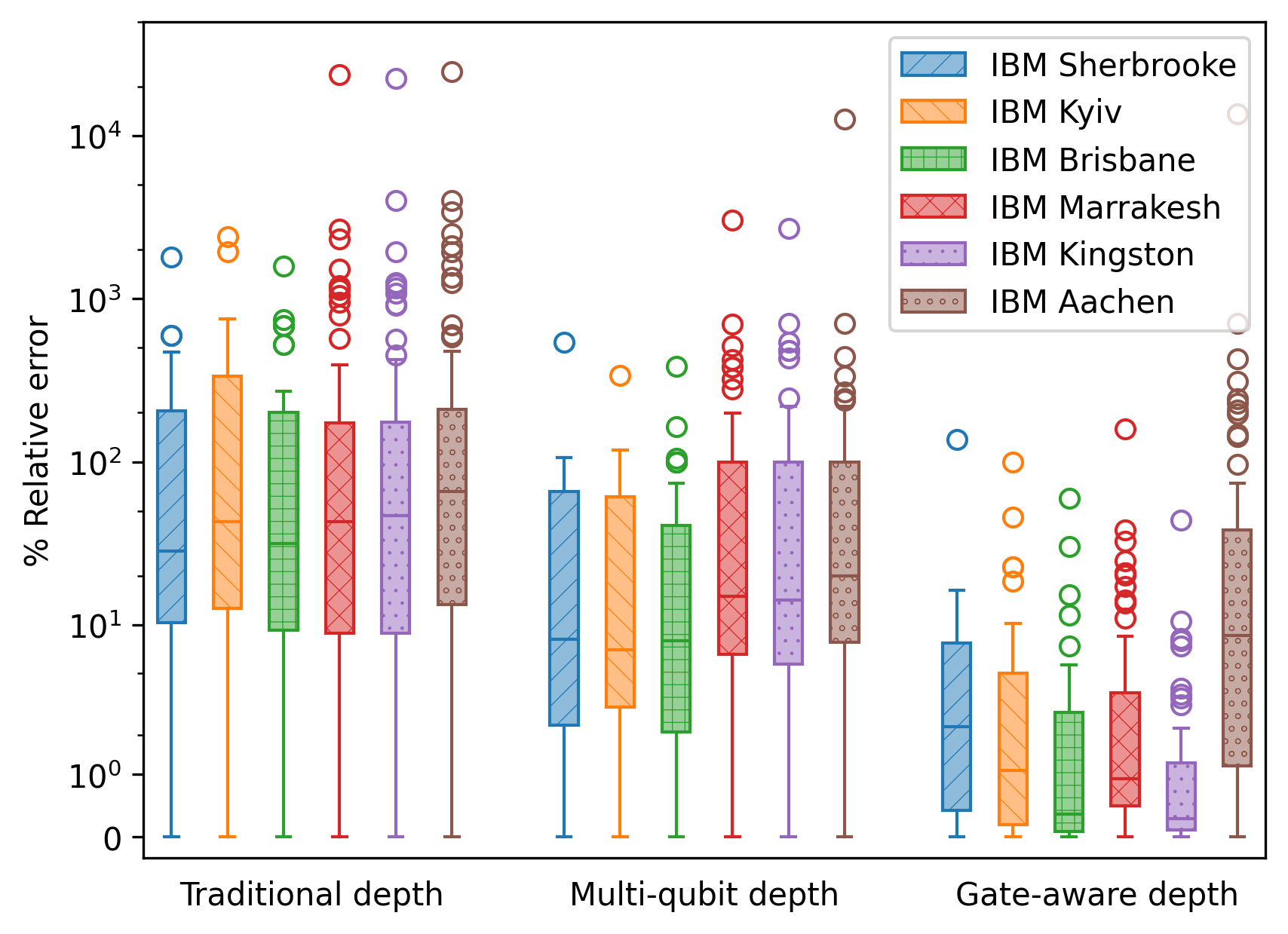}}
            \caption{Distribution of \%REs for predictions of relative runtime changes.}
            \label{fig:accuracy-rel-diff}
        \end{figure}

        Fig.~\ref{fig:accuracy-rel-diff} shows the \%RE distributions by metric. On all devices, gate-aware depth produced the lowest median \%RE, followed by multi-qubit depth with the second-lowest and traditional depth with the highest. On average across all devices, gate-aware depth reduced the median \%RE by 64  and 18 times relative to traditional and multi-qubit depth, respectively.

        For traditional depth, the third quartile exceeds 100\% RE on all devices. This means that, for at least one in four runtime predictions, the runtime change predicted by traditional depth differs from the true change by more than size of the true change itself. Multi-qubit depth performed only slightly better, with an average third quartile of 77.9\% RE. In comparison, gate-aware depth's third quartile was below 10\% RE for five out of six devices, only excluding the IBM Aachen. For those five devices, this means that at least three out of four predicted runtime differences deviated from the true runtime difference by less than 10\%. Only gate-aware depth's outliers exceeded the 100\% RE threshold regularly crossed by traditional depth.

        Gate-aware depth reduced the median \%RE and \%RE interquartile range less for the IBM Aachen, which occurs because this device has greater variability in gate execution times. The variation makes runtimes more sensitive to the circuit's placement on the physical hardware, which gate-aware depth does not account for.

    \subsection{Accuracy for Identifying Runtime-Optimal Circuit Versions}
        To compare the metrics' accuracies for identifying runtime-optimal circuit versions, we made 15 identifications using each metric on every device, one for each base circuit in the test suite.

\begin{figure}[tbp]
            \centerline{\includegraphics[width=0.45\textwidth]{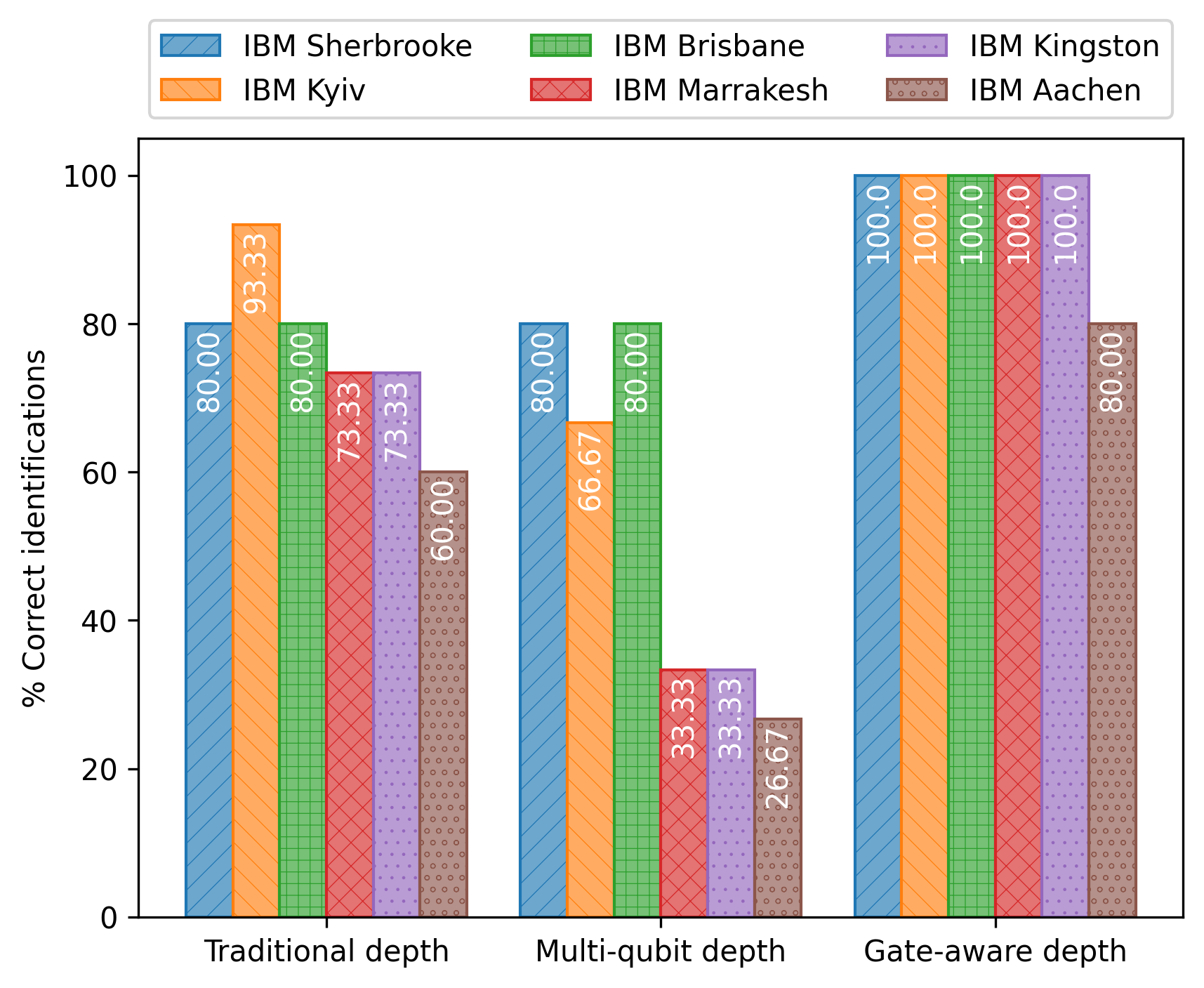}}
            \caption{Percentage of circuits with shortest-runtime version correctly identified.}
            \label{fig:accuracy-shortest}
        \end{figure}

        Fig.~\ref{fig:accuracy-shortest} shows the percentage of correct identifications by metric. Gate-aware depth made the most correct identifications for all devices, followed by traditional depth with the second-most and multi-qubit depth with the fewest (or tied for second). On average, gate-aware depth increased the number of correct comparisons by 20 and 43 percentage points over traditional and multi-qubit depth, respectively. In five out of six devices, gate-aware depth achieved a perfect 100\% accuracy rate, which no other metric achieved.
        
        Despite having the second-highest accuracy in predicting the relative difference in runtime (see Section~\ref{sec:accuracy-rel-diff}), multi-qubit depth had the lowest accuracy rate for identifying runtime-optimal circuit versions. A closer look at the Marrakesh identifications revealed that, for 8 out of the 10 incorrect identifications, the runtime-optimal compiled version tied with other versions for the lowest depth, which counts as incorrect. Ties are more likely for multi-qubit depth than the other metrics because it discards all information about single-qubit gates, eliminating a potential source of differentiation.

\section{Discussion}

    \subsection{Weight Selection}
        We verified that the weight selection method given by \eqref{eq:weight-config} chooses good values by varying the weight maps manually and checking the resulting accuracy for runtime predictions. Since the main difference between the weight maps in Table~\ref{tbl:avg-weight-maps} are the non-$RZ$ single-qubit gate weights, we parameterized them by \( w_s \) and tested values from \( 0 \) to \( 1 \) in increments of \( 0.01 \). The resulting weight maps are given in Table~\ref{tbl:manual-weight-maps}.
        
        \begin{table}[htbp]
            \begin{center}
                \caption{Manual Verification Weight Maps}
                \label{tbl:manual-weight-maps}
                \begin{tabular}{c||c|c|c|c|c}
                    $G$ & $ECR$ & $CZ$ & $RZ$ & $SX$ & $X$ \\
                    \hline
                    \( W'_{\text{Eagle}}(G) \) & $1.000$ & n/a & $0.000$ & $w_s$ & $w_s$ \\
                    \( W'_{\text{Heron}}(G) \) & n/a & $1.000$ & $0.000$ & $w_s$ & $w_s$ \\
                \end{tabular}
            \end{center}
        \end{table}
        
\begin{figure}[tbp]
            \centerline{\includegraphics[width=0.5\textwidth]{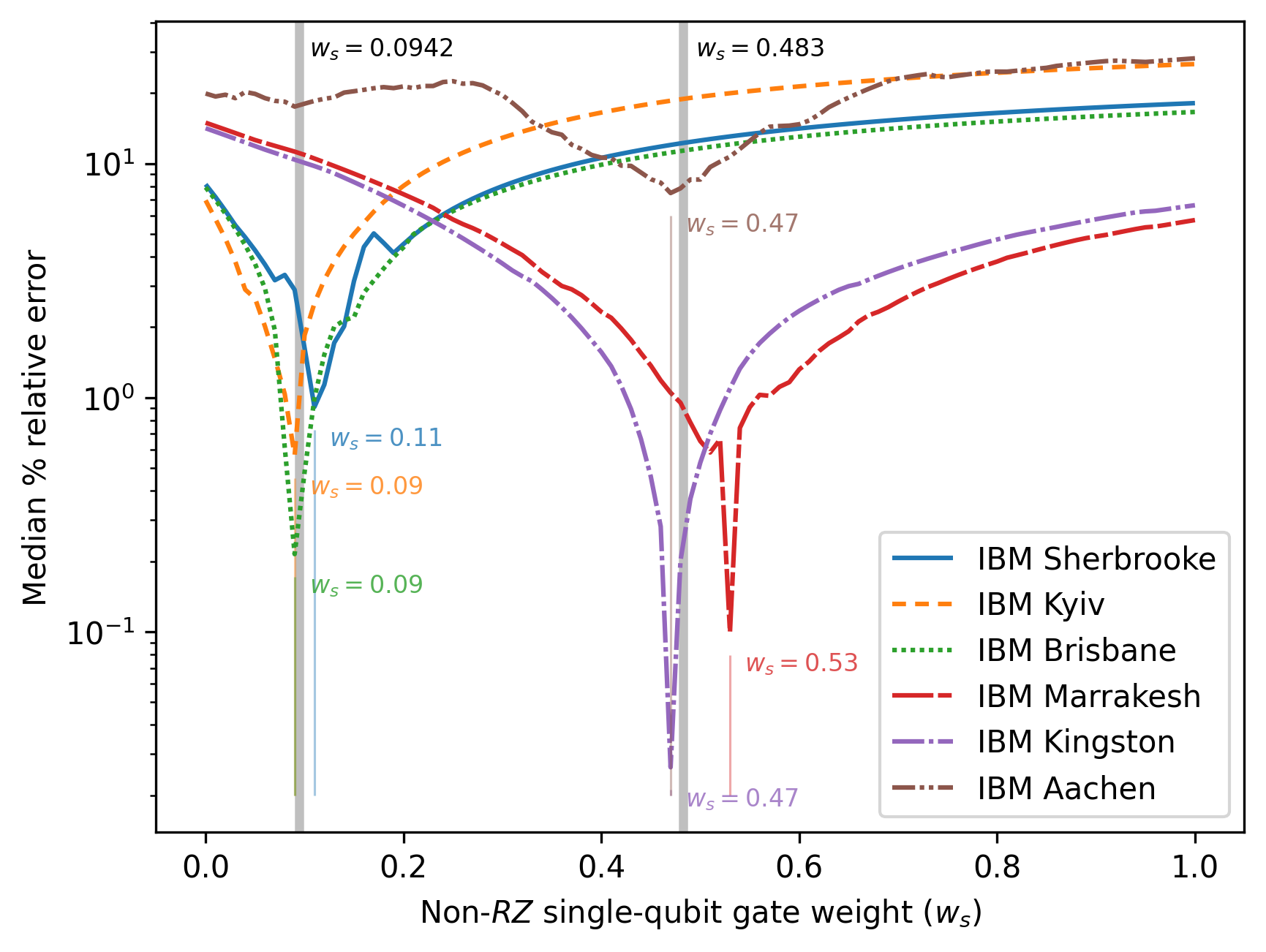}}
            \caption{Gate-aware depth's median \%RE for different values of non-\(RZ\) single-qubit gate weights \( w_s \). The thin vertical lines show the value of \( w_s \) that minimize median \%RE for a given device, and the thick ones show the real values calculated for the Eagle and Heron architectures in Section~\ref{sec:weight-maps}.}
            \label{fig:weight-dependence}
        \end{figure}

        Fig.~\ref{fig:weight-dependence} plots gate-aware depth's median \%RE for predicting relative differences in runtime against the weight \( w_s \). As expected, the values of \( w_s \) producing the minimum median \%RE for a given device cluster by architecture, with Eagle and Heron devices having optimal weights around 0.1 and 0.5, respectively. The true weights calculated in Section~\ref{sec:weight-maps}, shown by the thick vertical grey lines, fall in the middle of the cluster corresponding to their architecture. This shows that the weights given by \eqref{eq:weight-config} accurately capture gate time characteristics that are shared by devices of the same architecture.
        
    \subsection{Application to Compiler Comparison}        
        Researchers often compare compiler algorithms by compiling the same suite of test circuits using each algorithm and comparing the depths of the corresponding circuits, with greater relative reductions in depth indicating greater optimization ability. However, our results demonstrate that the two most commonly-used metrics, traditional and multi-qubit depth, are relatively inaccurate proxies for the true objective they aim to represent, runtime. Consequently, they provide inaccurate assessments of the relative performance of compilers. In comparison, gate-aware depth provides greater accuracy while still maintaining portability across devices of the same architecture.

        If inter-architecture portability is required, our results show that traditional depth is better able to identify runtime-optimal circuit versions because it accounts for all gates, while multi-qubit depth better predicts relative differences in runtime because two-qubit gates dominate circuit execution times in today's devices.

    \subsection{Non-Superconducting Devices}
        Although gate-aware depth is highly successful for the superconducting quantum devices tested, every qubit technology introduces changes in the process of scheduling and running quantum circuits that may affect its performance. For example, trapped-ion devices require additional shuttling time to shift ions around the trap, an operation which the circuit model fails to capture~\cite{schoenberger-2024-shuttling, bach-2025-efficient-shuttle}. We limited our experiment to IBM's superconducting devices because they offered direct access to circuit scheduling and runtime, but future works could extend the approach to other devices and technologies.

\section{Conclusion}

    In this paper, we show that circuit depth is inaccurate for comparing quantum circuit runtimes and propose a new metric that increases accuracy while maintaining hardware-agnosticism for devices of the same architecture. To do so, we identified variation in gate execution times as an underlying cause of the accuracy-portability tradeoff, and, in response, designed our metric to use average gate times for a given architecture. This approach achieves a middle-ground between the high portability of circuit depth and the high accuracy of circuit scheduling, thereby filling a gap in existing runtime comparison methods. We discuss the application of these findings to quantum compilation, and finally provide weight configurations for use with the current IBM Eagle and Heron architectures.

\section*{Acknowledgments}

    This material is based upon work supported by the DOE-SC Office of Advanced Scientific Computing Research MACH-Q project under contract number DE-AC02-06CH11357. This research used resources of the Oak Ridge Leadership Computing Facility, which is a DOE Office of Science User Facility supported under Contract DE-AC05-00OR22725.

\bibliography{refs} 

\vfill
\framebox{\parbox{.90\linewidth}{\scriptsize The submitted manuscript has been
created by UChicago Argonne, LLC, Operator of Argonne National Laboratory
(``Argonne''). Argonne, a U.S.\ Department of Energy Office of Science
laboratory, is operated under Contract No.\ DE-AC02-06CH11357.  The U.S.\
Government retains for itself, and others acting on its behalf, a paid-up
nonexclusive, irrevocable worldwide license in said article to reproduce,
prepare derivative works, distribute copies to the public, and perform publicly
and display publicly, by or on behalf of the Government.  The Department of
Energy will provide public access to these results of federally sponsored
research in accordance with the DOE Public Access Plan
\url{http://energy.gov/downloads/doe-public-access-plan}.}}

\end{document}